\documentclass[a4paper,twoside]{article}
\usepackage{bbm}
\usepackage{amsmath} 
\usepackage{graphics,graphicx}
\usepackage{verbatim}
\catcode`\@=11
\long\def\@makefntext#1{
\protect\noindent \hbox to 3.2pt {\hskip-.9pt  
$^{{\eightrm\@thefnmark}}$\hfil}#1\hfill}		

\def\@makefnmark{\hbox to 0pt{$^{\@thefnmark}$\hss}}	
	
\def\ps@myheadings{\let\@mkboth\@gobbletwo
\def\@oddhead{\hbox{}
\rightmark\hfil\eightrm\thepage}   
\def\@oddfoot{}\def\@evenhead{\eightrm\thepage\hfil
\leftmark\hbox{}}\def\@evenfoot{}
\def\sectionmark##1{}\def\subsectionmark##1{}}

\oddsidemargin=\evensidemargin
\newcounter{sectionc}\newcounter{subsectionc}\newcounter{subsubsectionc}
\renewcommand{\section}[1] {\vspace{12pt}\addtocounter{sectionc}{1} 
\setcounter{subsectionc}{0}\setcounter{subsubsectionc}{0}\noindent 
	{\tenbf\thesectionc. #1}\par\vspace{5pt}}
\renewcommand{\subsection}[1] {\vspace{12pt}\addtocounter{subsectionc}{1} 
\setcounter{subsubsectionc}{0}\noindent 
{\bf\thesectionc.\thesubsectionc. {\kern1pt \bfit #1}}\par\vspace{5pt}}
\renewcommand{\subsubsection}[1] {\vspace{12pt}\addtocounter{subsubsectionc}{1}
	\noindent{\tenrm\thesectionc.\thesubsectionc.\thesubsubsectionc.
	{\kern1pt \tenit #1}}\par\vspace{5pt}}

\newcounter{appendixc}
\newcounter{subappendixc}[appendixc]
\newcounter{subsubappendixc}[subappendixc]
\renewcommand{\thesubappendixc}{\Alph{appendixc}.\arabic{subappendixc}}
\renewcommand{\thesubsubappendixc}
	{\Alph{appendixc}.\arabic{subappendixc}.\arabic{subsubappendixc}}

\renewcommand{\appendix}[1] {\vspace{12pt}
        \refstepcounter{appendixc}
        \setcounter{figure}{0}
        \setcounter{table}{0}
        \setcounter{lemma}{0}
        \setcounter{theorem}{0}
        \setcounter{corollary}{0}
        \setcounter{definition}{0}
        \setcounter{equation}{0}
        \renewcommand{\thefigure}{\Alph{appendixc}.\arabic{figure}}
        \renewcommand{\thetable}{\Alph{appendixc}.\arabic{table}}
        \renewcommand{\theappendixc}{\Alph{appendixc}}
        \renewcommand{\thelemma}{\Alph{appendixc}.\arabic{lemma}}
        \renewcommand{\thetheorem}{\Alph{appendixc}.\arabic{theorem}}
        \renewcommand{\thedefinition}{\Alph{appendixc}.\arabic{definition}}
        \renewcommand{\thecorollary}{\Alph{appendixc}.\arabic{corollary}}
        \renewcommand{\theequation}{\Alph{appendixc}.\arabic{equation}}
        \noindent{\tenbf Appendix \theappendixc #1}\par\vspace{5pt}}
\newcommand{\subappendix}[1] {\vspace{12pt}
        \refstepcounter{subappendixc}
        \noindent{\bf Appendix \thesubappendixc. {\kern1pt \bfit #1}}
	\par\vspace{5pt}}
\newcommand{\subsubappendix}[1] {\vspace{12pt}
        \refstepcounter{subsubappendixc}
        \noindent{\rm Appendix \thesubsubappendixc. {\kern1pt \tenit #1}}
	\par\vspace{5pt}}

\newcommand{\textlineskip}{\baselineskip=13pt}
\newcommand{\smalllineskip}{\baselineskip=10pt}

\newcommand{\copyrightheading}[1]
	{\vspace*{-2.5cm}\smalllineskip{\flushleft
	 }}

\def\abstracts#1#2#3{{
	\centering{\begin{minipage}{4.5in}\footnotesize\baselineskip=10pt
	\parindent=0pt #1\par 
	\parindent=15pt #2\par
	\parindent=15pt #3
	\end{minipage}}\par}} 

\def\keywords#1{{
	\centering{\begin{minipage}{4.5in}\footnotesize\baselineskip=10pt
	{\footnotesize\it Keywords}\/: #1
	 \end{minipage}}\par}}

\renewenvironment{thebibliography}[1]
        {\frenchspacing
	 \ninerm\baselineskip=11pt
         \begin{list}{\arabic{enumi}.}
        {\usecounter{enumi}\setlength{\parsep}{0pt}     
	 \setlength{\leftmargin 12.7pt}{\rightmargin 0pt}
         \setlength{\itemsep}{0pt} \settowidth
	{\labelwidth}{#1.}\sloppy}}{\end{list}}

\newcounter{itemlistc}
\newcounter{romanlistc}
\newcounter{alphlistc}
\newcounter{arabiclistc}

\def\pmb#1{\setbox0=\hbox{#1}
	\kern-.025em\copy0\kern-\wd0
	\kern.05em\copy0\kern-\wd0
	\kern-.025em\raise.0433em\box0}

\def\fnt#1#2{\footnotetext{\kern-.3em
	{$^{\mbox{\scriptsize #1}}$}{#2}}}

\def\fpage#1{\begingroup
\voffset=.3in
\thispagestyle{empty}\begin{table}[b]\centerline{\footnotesize #1}
	\end{table}\endgroup}

\def\runninghead#1#2{\pagestyle{myheadings}
\markboth{{\protect\footnotesize\it{\quad #1}}\hfill}
{\hfill{\protect\footnotesize\it{#2\quad}}}}
\font\tenrm=cmr10
\font\tenit=cmti10 
\font\tenbf=cmbx10
\font\bfit=cmbxti10 at 10pt
\font\ninerm=cmr9

\font\eightrm=cmr8

\newtheorem{theorem}{Theorem}

\newtheorem{lemma}{Lemma}

\newcommand{\proof}[1]{{\bf Proof.} #1 $\proofend$}

\def\qed{\hbox{${\vcenter{\vbox{	          
   \hrule height 0.4pt\hbox{\vrule width 0.4pt height 6pt
   \kern5pt\vrule width 0.4pt}\hrule height 0.4pt}}}$}}

\newcommand{\proofend}{\hfill\rule{2mm}{2mm}\medskip}
\newcommand{\Eqref}[1]{Eq.~(\ref{#1})}

\begin{document}
\setlength{\textheight}{8.0truein}    

\runninghead{Entanglement transformations of pure Gaussian states}
            {authors}

\normalsize\textlineskip
\thispagestyle{empty}
\setcounter{page}{1}

\copyrightheading{}	

\vspace*{0.88truein}

\fpage{1}
\centerline{\bf
ENTANGLEMENT TRANSFORMATIONS OF PURE GAUSSIAN STATES}
\vspace*{0.37truein}
\centerline{\footnotesize 
G.\ Giedke$^{1,2}$, J.\ Eisert$^{3,4}$, J.I.\ Cirac$^{2}$, and M.B.\ Plenio$^4$}
\vspace*{0.015truein}
\centerline{\footnotesize\it $^{1}$ 
4.~Physikalisches Institut, Universit{\"a}t Stuttgart, Pfaffenwaldring
57, D-70550 Stuttgart, Germany}
\centerline{\footnotesize\it 
$^{2}$ Max-Planck--Institut f{\"u}r Quantenoptik,
Hans-Kopfermann-Stra{\ss}e, D-85748 Garching,
Germany}
\centerline{\footnotesize\it $^{3}$ 
Institut f{\"u}r Physik, Universit{\"a}t Potsdam, Am Neuen Palais 10, D-14469 Potsdam, Germany}
\centerline{\footnotesize\it $^{4}$ QOLS,
Blackett Laboratory, Imperial College London, SW7 2BZ London, UK}

\baselineskip=10pt
\vspace*{0.21truein}
\abstracts{
We present a theory of entanglement transformations of Gaussian pure
states with local Gaussian operations and classical communication.
This is the experimentally accessible set of operations that can be
realized with optical elements such as beam splitters, phase shifts
and squeezers, together with homodyne measurements.  We provide a
simple necessary and sufficient condition for the possibility to
transform a pure bipartite Gaussian state into another one. We
contrast our criterion with what is possible if general local
operations are available.
}{}{}
\vspace*{10pt}
\vspace*{3pt}

\vspace*{1pt}\textlineskip	


\section{Introduction}

For optical systems, both reliable sources producing Gaussian quantum
states and efficient detection schemes such as homodyne detection are
experimentally readily available \cite{Vogel}. Quantum states can also
be manipulated in an accurate manner by means of optical elements such
as beam splitters and phase plates. In fact, it has been realized that
such systems with canonical variables -- often referred to as
continuous-variable systems -- in Gaussian quantum states offer a
promising potential for realistic quantum information
processing. The so-called teleportation schemes for continuous-variables
have been theoretically proposed and experimentally implemented
\cite{Tel}, generation of entanglement has been studied \cite{App},
and cloning \cite{Clo}, Bell-type schemes \cite{Bell}
and cryptographic protocols \cite{Crypt} have been suggested, to name a few. 
In addition to the work on purely
optical systems, continuous atomic variables have been investigated in
great detail, e.g., when studying collective spin states of a
macroscopic sample of atoms \cite{Polzik}. From the perspective of
the theory of quantum entanglement, the concepts of separability
\cite{Sep} and distillability \cite{Dist}, as well as entanglement
quantification \cite{CanQuant} have been extended to systems with
canonical variables. All these investigations complement the original
studies of entanglement in quantum information science in the
finite-dimensional regime.

In the light of these successes it would be desirable to have
tools at hand that help with deciding whether an envisioned task can
be achieved in a feasible manner or not. One would then ask for
mathematical criteria whether a certain state transformation can be
performed under a class of quantum operations that reflects the
natural physical constraints of a given set-up. Such a tool proved
very useful in the finite-dimensional setting. Often referred to as
majorization criterion \cite{Pure}, it is a criterion for the set of
local operations with classical communication (LOCC), which is the
natural choice for finite-dimensional bi-partite systems: A pure state
of a bi-partite system can be transformed into another pure state
under LOCC if and only if the reductions of the state are finally more
mixed than initially. The term `more mixed' has to be understood in
the sense of majorization theory. Hence, the problem of deciding
whether a particular transformation can be performed in principle --
which can be an extremely difficult task -- can be linked to a simple
majorization relation.

A first step towards a theory of entanglement transformations for
bi-partite systems with canonical variables has been undertaken in
Ref.\ \cite{MixedCrit}: This criterion can in fact be applied to mixed
Gaussian states of two modes, and the set of operations is a
practically important set of feasible Gaussian operations. It does
not, however, include measurements and classical communication.  For
pure states, in contrast, one could hope for a general criterion of
entanglement transformation under all Gaussian local operations with
classical communication (GLOCC). Gaussian operations are those
operations that preserve the Gaussian character of states, and
correspond exactly to those operations that can be implemented by
means of optical elements such as beam splitters, phase shifts and
squeezers together with homodyne measurements -- all operations that
are experimentally accessible with present technology
\cite{Operation3,Operation1,Operation2}. A complete characterization
of Gaussian operations has been presented in Refs.\
\cite{Operation3,Operation2}.

\begin{figure}[t]
\centerline{
\includegraphics[width=83mm]{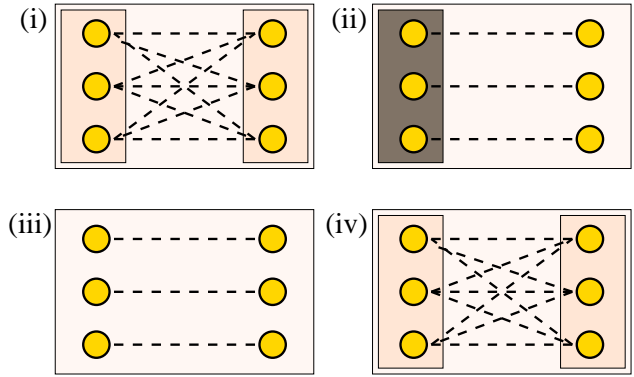}
} \vspace{.2cm} \caption{Any transformation from one pure Gaussian
state of $n\times n$ modes to another such state by means of local
Gaussian operations with classical communication can be decomposed
into three steps: (i) First, by means of local unitary Gaussian
operations the state can be transformed into a product of $n$ two-mode
squeezed vacua as depicted in (ii). (iii) By means of (measuring)
local Gaussian operations, classical communication of the outcomes and
appropriate local Gaussian unitary operations one obtains a different
product of two-mode squeezed vacua. (iv) The last step is to apply
local Gaussian unitary operations in order to obtain the desired final
state.}
\end{figure}

In this paper we show that such a general criterion for pure-state
Gaussian entanglement transformations can in fact be formulated: We
present a necessary and sufficient criterion for pure-state
entanglement transformations of bi-partite $n\times n$-mode systems
under GLOCC. The criterion itself will turn out to be very simple:
after having shown that each pure Gaussian state of an $n\times n$
mode system is equivalent up to local Gaussian unitary operations to
two-mode squeezed states (see also Ref.\ \cite{Botero}), it will turn
out that the criterion merely amounts to an element-wise comparison of
squeezing vectors.

\section{Notation}

We will consider general Gaussian states $\rho$ of $n\times n$
field modes, associated with the $4n$ canonical coordinates
$R:=(X_{1},P_{1},\ldots,X_{2n},P_{2n})$. A {\em Gaussian state} is
a state $\rho$ the {\em characteristic function}
\begin{eqnarray}
\chi_{\rho}(\xi)= \text{tr}[\rho W_\xi]
\end{eqnarray} 
of which is a  Gaussian
function in phase space, where $W_\xi:=\exp(-i \xi^{T} R)$ is the
{\em Weyl (displacement) operator}. The state $\rho$ can then be
written as
\begin{eqnarray}
    \rho=\pi^{-2n}\int_{{\mathbbm{R}}^{4n}} d\xi \exp( - \frac{1}{4}\xi^{T}
    \gamma \xi + i d^{T}\xi) W_\xi,
\end{eqnarray}
where $\gamma=\gamma^{T}$ is
the {\em covariance matrix} (CM)
incorporating the second moments, and $d$ is the vector of the first
moments. The first moments can be made to vanish via local operations
and contain no information about entanglement. Only the second 
moments will therefore be of interest in the subsequent analysis. For example,
the CM of a {\em two-mode squeezed state} $\rho_r$
is given by
\begin{equation}\label{CMtmss}
    \gamma_{r}=\left(
    \begin{array}{cc}
    A_{r} & C_{r}\\
    C_{r}& A_{r}
    \end{array}
    \right),
\end{equation}
$A_{r}:=\cosh(r){\mathbbm{1}}$ and    $C_{r}:=
\sinh(r)\Lambda$, with $\Lambda = \text{diag}(1,-1)$,  
where $r\in[0,\infty)$ is the (two-mode) {\em squeezing parameter}. The
canonical commutation relations can be formulated as
$[R_{j},R_{k}]= i \sigma_{j,k}$, $j,k=1,\ldots,4n$, with
\begin{eqnarray}
    \sigma=\bigoplus\limits_{i=1}^{2n} \sigma_{1}, \,\,\,\sigma_{1}=
    \left(
    \begin{array}{cc}
    0 & -1 \\
    1 & 0
    \end{array}
    \right)
\end{eqnarray}
being the {\em symplectic matrix}. 
In standard matrix theory the unitary diagonalization of a matrix
and the resulting eigenvalues are key concepts. In the study of
Gaussian states and their covariance matrices this role is taken
by the idea of symplectic diagonalization. Any covariance matrix
$\gamma$ can be transformed into a diagonal matrix 
$S\gamma S^T$ by some symplectic transformation $S$.
The diagonal elements of $S\gamma S^T$ form the
symplectic spectrum. The {\em symplectic spectrum}
of $\gamma$ can be directly calculated as the modulus of the eigenvalues
of $\sigma \gamma$.
The {\em Heisenberg uncertainty principle} can be written
as $\gamma\geq i \sigma$ \cite{Symplectic}, which is also necessary
and sufficient for the real, positive matrix $\gamma$ to be a CM.

{\em Gaussian operations}
\cite{MixedCrit,Operation3,Operation1,Operation2,Operation0} are
those quantum operations (completely positive maps) that map all
Gaussian states on Gaussian states. The practically most important
subset is the set of Gaussian unitary operations, which are
reflected by so-called {\em symplectic transformations}
\begin{equation}
\gamma\longmapsto S \gamma S^{T}
\end{equation} 
with $S\sigma S^{T}=\sigma$ on
the level of covariance matrices. These are those unitary
operations that can be realized by means of beam splitters,
squeezers, and phase shifts. The most general ``pure'' Gaussian operation
can be conceived as a concatenation of Gaussian unitary operations
(on possibly a larger set of modes), together with homodyne
measurements \cite{Operation3,Operation1,Operation2}. In the
present context, it is most convenient to employ the
isomorphism between completely positive maps and states
\cite{Jam}, as explicitly analyzed and completely characterized in
Ref.\ \cite{Operation3}. A general Gaussian map gives rise to a
transformation $\gamma\longmapsto \gamma'$, where $\gamma'$
is a Schur complement \cite{Horn}
of the form
\begin{eqnarray}\label{gmap}
    \gamma' = \tilde \Gamma_{1}-\tilde\Gamma_{12} (
    \tilde \Gamma_{2}+\gamma
     )^{-1}\tilde \Gamma_{12}^{T}.
\end{eqnarray}
The CM
\begin{eqnarray}
    \Gamma=\left(
    \begin{array}{cc}
    \Gamma_{1} & \Gamma_{12}\\
    \Gamma_{12}^T & \Gamma_{2}
    \end{array}
    \right)
\end{eqnarray}
defined on $4n$ modes specifies the actual Gaussian operation, where
\begin{eqnarray}
\tilde\Gamma :=(1\oplus \Lambda) \Gamma (1\oplus \Lambda),
\end{eqnarray}
and
$\Lambda:= \text{diag}(1,-1,1,-1,\ldots,1,-1)$.

\section{A general criterion}
Before we are in the position to state the theorem, we introduce a
particularly useful normal form of Gaussian pure states of $n\times n$
modes.  This normal form has also been found independently by Botero
and Reznik \cite{Botero}. For reasons of completeness of this paper,
we will however present an alternative proof in the notation used here
in Appendix A.

\begin{lemma}[Standard form of pure $n\times n$-states]\label{Lstdform} 
Any pure Gaussian state $\rho$ of $n\times n$ modes can be transformed
by local unitary Gaussian operations into a state which is a
tensor product of $n$ pure two-mode squeezed states (TMSS) with
squeezing parameters $r_{1}\geq \ldots\geq r_{n}\geq 0$. 
\end{lemma}

This means that in any orbit with respect to local Gaussian
unitary operations there is a product state of $n$ TMSS. 
When considering entanglement transformations
with two pure Gaussian states $\rho$ and $\rho'$
of $n\times n$
modes, we can consider them without loss of generality to be of this
normal form, characterized by vectors
$r$ and $r'$ of ascendingly ordered squeezing parameters,
respectively. We say
that 
\begin{equation}
	r\geq r' 
\end{equation}
iff
    $r_{k}\geq r_{k}'$
for all $k=1,\ldots,n$, which allows the concise statement of our
    main result announced before.

\begin{theorem}[Pure-state entanglement transformations]\label{maintheorem}
Let $\rho$ and $\rho'$ be two $n\times n$ pure states with CM $\gamma$
and $\gamma'$, characterized by ordered squeezing vectors $r$ and
$r'$, respectively. Then $\rho$ can be transformed into $\rho'$ be
local Gaussian operations with classical communication iff $r\geq
r'$, abbreviated as
\begin{equation}\label{MainCrit}
    \rho\longrightarrow \rho'\,\text{ under GLOCC,\, iff }r\geq r'.
\end{equation}
\end{theorem} 

The subsequent Lemmas prepare the proof. Lemma~\ref{Lsympl} is a tool that
connects a relation between two positive symmetric matrices to a relation
involving the symplectic spectrum of the matrices. 

\begin{lemma}[Symplectic spectrum]\label{Lsympl}
Consider real positive $2n\times 2n$ matrices $M_1,M_2\geq 0$. Let
$s(M_1)$ and $s(M_2)$ 
be the vectors consisting of the (ascendingly ordered) symplectic
eigenvalues of $M_1$ and $M_2$, respectively. Then
\begin{eqnarray}\label{eq2}
     M_1\geq M_2 \Longrightarrow s(M_1)_{k}\geq s(M_2)_{k}\,\,\,\,\,\forall
     \,k=1,\ldots,n.
\end{eqnarray}
\end{lemma}
\proof{As a consequence of  $M_1\geq M_2$ and
that $\sigma $ is a skew symmetric 
matrix we find $-\sigma M_1\sigma
\geq -\sigma M_2\sigma $. As $M_1$ is a strictly positive real
symmetric matrix, $M_1^{1/2}$ is well-defined and symmetric, and
hence,
\begin{eqnarray}
     -M_1^{1/2} \sigma  M_1 \sigma  M_1^{1/2}\geq -M_1^{1/2} \sigma  M_2 \sigma  M_1^{1/2},
\end{eqnarray}
and likewise 
$     -M_2^{1/2} \sigma  M_1 \sigma  M_2^{1/2}\geq -M_2^{1/2} \sigma  
	M_2 \sigma  M_2^{1/2}$.
  From the corollary of Weyl's theorem known as
monotonicity theorem \cite{Horn} it follows that  
\begin{eqnarray}
     \lambda_{k}(-M_j^{1/2} \sigma  M_1 \sigma  M_j^{1/2})\geq \lambda_{k}(-M_j^{1/2} \sigma  M_2 \sigma
     M_j^{1/2})
\end{eqnarray}
for all $j=1,2$ and  $k=1,\ldots,n$, where
$\lambda(M)$ for any real symmetric matrix $M$ is the vector of
(ascendingly ordered) eigenvalues. But
\begin{eqnarray}
\lambda_{k}(-M_1^{1/2}\sigma M_2\sigma M_1^{1/2})& =&
\lambda_{k}(-M_2^{1/2}\sigma M_1\sigma M_2^{1/2})\nonumber\\
 &\geq& \lambda_{k}(-M_2^{1/2}\sigma M_2\sigma M_2^{1/2}). 
\end{eqnarray}
Hence,
\begin{eqnarray}
     \lambda_{k}(-M_1\sigma M_1\sigma)\geq \lambda_{k}(-M_2\sigma M_2\sigma )
\end{eqnarray}
for all $k=1,\ldots,n$, which implies the validity
of the right hand side of Eq.\ (\ref{eq2}).}

Lemma~\ref{Loneloc} shows that we can restrict our considerations to the
class of one-local pure transformations. These are then shown in
Lemma~\ref{Lunilat} to have a simple form from which the desired
result can be read off.

First, note that it does not restrict generality to
impose the condition that in the course of the protocol the joint
state is pure at all stages.  This means that the CM associated with
the completely positive map realizing the protocol can be taken to be
of direct sum form.  Moreover, as in the finite-dimensional case, one
does not have to consider all local operations with classical
operations, but only those with one-way classical communication:

\begin{lemma}[One-local operations are sufficient]\label{Loneloc}
Given two $n\times n$ pure states $\rho$ and $\rho'$
with CM $\gamma$ and $\gamma'$, respectively.
Then $\rho$ can be transformed into $\rho'$ under GLOCC,
iff $\rho$ can be transformed into $\rho'$ by one-local Gaussian
operations (i.e., Gaussian local operations in system $A$
with communication from system $A$ to $B$ only together with local
Gaussian unitary operations in $B$).
\end{lemma}

\proof{The analogous statement in the finite-dimensional setting
has been proven in Ref.\ \cite{Lo}: For general entanglement
transformations of finite-dimensional pure states it does not
restrict generality to make use of one-local operations
only. It remains to be shown that the construction of
Ref.\ \cite{Lo} can be performed in the
infinite-dimensional setting for Gaussian operations:
Every GLOCC that is associated with a pure Gaussian state
can be conceived as a sequence of elementary steps. Each such
elementary step consists of local Gaussian unitary operations
in one system, local Gaussian measurements, and
the communication of the classical outcomes. 
Hence, it only has to be shown that 
each Gaussian measurement in system $B$ can be 
equivalently implemented by means of a 
Gaussian measurement
in system $A$, accompanied by appropriate local Gaussian unitary
operations in both $A$ and $B$. To see that this is the case, note 
firstly that the Schmidt decomposition can be applied
in this infinite-dimensional case. Secondly, the unitary operation
mapping any pure Gaussian state 
onto its Schmidt decomposition is
a Gaussian unitary operation, 
as can be inferred from Lemma~\ref{Lstdform}. According to the
argument of Ref.\ \cite{Lo},  therefore, it follows that
for any Gaussian state vector $|\omega\rangle_{B'}$, any bi-partite
Gaussian state vector $|\psi\rangle_{AB}$, any unitary
$U_{BB'}$ corresponding to a Gaussian unitary transformation, and any Gaussian
state vector $|\phi\rangle_{B'}$ (potentially corresponding to an improper
state \cite{improper}), there exist unitaries $V_{AA'}$ and $V_B$ such that 
\begin{eqnarray}
	\langle\phi|_{B'}({\mathbbm{1}}\otimes U_{BB'}) |\psi\rangle_{AB}|\omega\rangle_{B'}=
	\langle\phi|_{A'} (V_{A A'}\otimes V_B ) |\psi\rangle_{AB}|\omega\rangle_{A'}.
\end{eqnarray}
The unitaries $ V_{AA'}$ and $V_B$
in turn also correspond to Gaussian unitary
operations. Therefore, any Gaussian measurement in system $B$ leads to
the same final pure state as a Gaussian measurement in system $A$,
followed by appropriate local Gaussian unitary operations in both parts.}

Ref.\ \cite{Operation3} gives the general form of a Gaussian
local operation with classical communication. It is considerably
simplified for the special case that a local operation is 
implemented in one of the two parts of the joint system only. Then
Eq.\ (\ref{gmap}) becomes

\begin{lemma}[Unilateral transformations]\label{Lunilat}
Let $\gamma$ be a CM of a Gaussian state of an $n\times n$ mode system
consisting of systems $A$ and $B$, partitioned as
\begin{equation}\label{np}
	\gamma=\left(\begin{array}{cc}
	A & C \\
	C^T & B
	\end{array}
	\right).
\end{equation}
The
CM $\gamma'$ after application of a general
Gaussian local operation in system $A$ characterized by a CM
\begin{eqnarray}
    \Gamma_A=\left(
    \begin{array}{cc}
        \Gamma_{1A} & \Gamma_{12A}\\
        \Gamma_{12A}^T & \Gamma_{2A}
    \end{array}\right)
\end{eqnarray}
 is given by a matrix of
the form as in Eq.\ (\ref{np}) with
\begin{subequations}
\begin{eqnarray}
    A'&=&\Gamma_{1A} - \tilde
    \Gamma_{12A}(\tilde\Gamma_{2A}+A)^{-1}
    \tilde\Gamma_{12A}^{T},\\
    B'&=& B- C^T (\tilde \Gamma_{2A}+A)^{-1} C,\\
    C'&=&\tilde \Gamma_{12A}(\tilde\Gamma_{2A}+A)^{-1} C.
\end{eqnarray}
\end{subequations}
\end{lemma}
As an example, we now discuss the unilateral transformation that
transforms a TMSS with squeezing parameter $r$ into another TMSS with
squeezing parameter $r'<r$.  On the level of covariance matrices, this
transformation can be achieved by applying quantum operations in
system $A$ only. The Gaussian operation that realizes this map is
associated with a $4\times 4$ CM $\Gamma_A$, which is given by
\begin{equation}
	\Gamma_A= \left(
	\begin{array}{cc}
	A_{r''} & C_{r''}\\
	C_{r''}  & 	A_{r''}
	\end{array}
	\right),
\end{equation}
where as before $A_{r''} = \cosh(r''){\mathbbm{1}}$ and
$C_{r''} = \sinh(r'') 
\Lambda$. The squeezing parameter
$r''\in[0,\infty)$ is defined via
\begin{eqnarray}
    \cosh(r'') = \frac{\cosh(r)\cosh(r') -1}{\cosh(r) -
    \cosh(r')}.
\end{eqnarray}
Physically, this operation can be implemented in two steps: first, one
implements an appropriate Gaussian unitary operation on both the
single mode of system $A$ and an additional vacuum field mode. This
can be done by applying suitable linear optical elements. Then, in a
second step, one realizes a homodyne detection in the additional field
mode. When considering the second moments only as we have done
throughout the paper, no action is needed in system $B$. In fact, the
classical outcome needs to be communicated only to apply the
appropriate local displacement in phase space in system $B$.
Equipped with Lemmas~\ref{Lsympl}-\ref{Lunilat} we can now prove the
Theorem. 

\smallskip \textbf{Proof of the Theorem:} 
Let us without loss of generality assume that $\rho$ already is of the
normal form of Lemma~\ref{Lstdform}. Define for a squeezing vector $r$ the
$2n\times 2n$ matrices $c(r)$ and $s(r)$ as
\begin{eqnarray}
	c(r):= \bigoplus_k \cosh(r_k){\mathbbm{1}}_2,\,\,\,
	s(r):=  \bigoplus_k \sinh(r_k)  \Lambda.
\end{eqnarray}
According to Lemma~\ref{Lunilat} the general form of a CM after a
one-local Gaussian operation is given by $\gamma'$ partitioned as in
Eq.~(\ref{partit}), with
\begin{subequations}
\begin{eqnarray}
    A'&=& \Gamma_{1A}- \tilde \Gamma_{12A}[\tilde \Gamma_{2A}+
    c(r)]^{-1}\tilde \Gamma_{12A}^{T},\\
    B'&=& S^{T}\left( c(r) 
    - s(r) [\tilde \Gamma_{2A}+c(r)]^{-1} s(r)\right) S, \\
    C'&=& \tilde \Gamma_{12A}[\tilde \Gamma_{2A}+c(r)]^{-1}
    s(r) S,
\end{eqnarray}
\end{subequations}
where $S$ is the symplectic transformation corresponding to the
Gaussian unitary operation in system $B$. Again, the final state
can be taken to be of normal form. Imposing the condition
  $A'=B'=c(r')$   and
$C'=s(r')$ and solving for $\Gamma$ yields
\begin{subequations}
\begin{eqnarray}
    \Gamma_{1A}&=& s(r')
    \left(\tilde S^{T} [\tilde S^{T} c(r)\tilde S-c(r')]
    \tilde S\right)^{-1} s(r')+c(r') \\
    \Gamma_{12A}&=& s(r')\tilde S^{-1}
    \left[\tilde S^{T} c(r)\tilde S-c(r')\right]^{-1}
    s(r), \\
    \Gamma_{2A}&=& s(r) \left[\tilde S^{T} c(r)\tilde
    S-c(r')\right]^{-1}
    s(r) - c(r),
\end{eqnarray}
\end{subequations}
where $\tilde S$ is the symplectic
transformation $\tilde S:= \Lambda S \Lambda$. From the
expressions for $\Gamma_{1A}$ and $\Gamma_{2A}$ one finds that
$\Gamma_{A}\geq i\sigma $ can only hold if   
${\tilde
S}^{T}c(r){\tilde S}-c(r')\geq 0$,
meaning that 
\begin{eqnarray}
	s[{\tilde
S}^{T}c(r){\tilde S}] \geq s[c(r')],
\end{eqnarray}
for which by Lemma~\ref{Lsympl} it is necessary that $r\geq r'$.
On the other hand, if $r\geq r'$ we can choose $S={\mathbbm{1}}$
and one finds by direct calculation that $\Gamma_{A}$ then describes a
product of two-mode squeezed states in standard form with
squeezing vector $r''$; the $k$-th entry $r_{k}''$ is given by
\begin{eqnarray}
    \cosh(r_{k}'') = \frac{\cosh(r_k)\cosh(r'_k) -1}{\cosh(r_k) -
    \cosh(r'_k)}.
\end{eqnarray}
This argument shows that the conditions are also sufficient for
the   transformation of
the states under GLOCC.
\proofend

\section{Comparison with the majorization criterion}
The simplicity of the above criterion is quite astonishing, as
compared to the majorization structure in finite-dimensions
\cite{Pure}. As a corollary, it follows that not only pure-state
distillation is not possible, but in fact any pure-state collective
Gaussian quantum operation. In particular, so-called {\it catalysis of
entanglement manipulation}, as it has been studied in the
finite-dimensional case \cite{cata}, can not occur: the metaphor of
catalysis refers to the effect that in finite dimensions, it can
happen that $\rho\not \longrightarrow \rho'$ under LOCC, but
\begin{equation}
	\rho\otimes
\omega  \longrightarrow \rho'\otimes \omega \text{ under LOCC}
\end{equation}
for some appropriate catalyst state $\omega$. It should moreover be
mentioned that -- as Gaussian transformations can be made
deterministic \cite{Operation2} -- there is no space for distinct
criteria for the stochastic interconversion between states which is
again in contrast to the finite dimensional case \cite{Vidal}, and
this case is also covered by the above Theorem.

To explore to what extent the restrictions of the
Ineq.~(\ref{MainCrit}) arise from the limitation to local
\emph{Gaussian} operations and which remain even if \emph{general}
local operations are allowed, we apply the 
state-transformation criterion of Nielsen \cite{Pure} to pure Gaussian
states for two typical examples. 
\Eqref{MainCrit} has two main features. First, it implies that with
Gaussian operations one
cannot ``concentrate'' two-mode squeezing, i.e., increase the largest
two-mode squeezing parameter that is available. This becomes particularly
clear when considering two-mode squeezed states $\rho_r$ with
CM $\gamma_r$, where $r\in [0,\infty)$. They correspond to state vectors
\begin{equation}
	|\psi_r\rangle = \cosh^{-1}(r/2) \sum_{k=0}^\infty
	\tanh^k(r/2)|k\rangle_A |k\rangle_B,
\end{equation}
where $\{|k\rangle: k\in{\mathbbm{N}}\}$ denotes the Fock basis. 
E.g., the transformation 
\begin{equation}\label{ex1}
\rho_r^{\otimes n} \longrightarrow \rho_s \otimes
\rho_0^{\otimes(n-1)}, \,\, r<s
\end{equation}
is not possible with GLOCC, 
no matter how close $s$ is to $r$ or how large $n$
is. Second, one cannot ``dilute'' two-mode squeezing., i.e., 
the final $r$-vector cannot have more non-zero entries (or indeed
entries strictly larger than any given threshold $s_0\geq0$)
than the initial one. In particular, it is impossible to locally
implement 
\begin{equation}\label{ex2}
\rho_s \otimes \rho_0 \longrightarrow \rho_r\otimes\rho_r,\,\, r>0
\end{equation}
with GLOCC no matter how large $s$ or how small $r$ are. 

We show now that transformations of the kind (\ref{ex1}) \emph{can} be
realized if \emph{arbitrary} local operations (LOCC) are allowed and
$r$ is sufficiently large.  This shows (not all too surprisingly) that
LOCC is more powerful than GLOCC, even when both initial and final
state are required to be Gaussian. On the other hand, we show that
(\ref{ex2}) is \emph{not possible} even with general (not necessarily
Gaussian) local operations accompanied with classical communication.

For arbitrary local operations the transformation properties between
bipartite pure states are governed by their Schmidt coefficients. 
The ordered list of 
Schmidt coefficients of $\rho_s $ is given by the vector $m$ with
components
\begin{equation}
	m_k=(1-\eta) \eta^k,
\end{equation}
where $\eta:= \tanh^2(s/2)$ and $k=0,1,...$. The ordered list of 
Schmidt coefficients of the
initial state $\rho_r\otimes \rho_r$ is the vector
\begin{equation}\label{Scoeff2tmss}
l:=(1-\lambda)^2 (1, 
	\lambda,  \lambda,
	 \lambda^2,\lambda^2 , \lambda^2,
	 \lambda^3, \lambda^3,
	 \lambda^3, \lambda^3,...
	),
\end{equation}
with $\lambda:=\tanh^2 (r/2)$. The transformation
$\rho_r\otimes \rho_r\longrightarrow \rho_s\otimes \rho_0$ under LOCC
is possible if and only if
\begin{equation}\label{cr}
	\sum_{k=0}^N l_k \leq 	\sum_{k=0}^N m_k
\end{equation}
for all $N=0,1,...$.  Obviously, if we allow for a summation over more
than the first $N$ positive terms on the left hand side of Ineq.\
(\ref{cr}) while still having the inequality satisfied, the
transformation is also possible. So, 
certainly $\rho_r\otimes \rho_r\longrightarrow \rho_s\otimes \rho_0$
 under LOCC holds if
\begin{eqnarray}\label{inj}
	(1-\lambda)^2 \sum_{k=0}^N (k+1) \lambda^k &=& 	(1-\lambda)^2 
	\frac{d}{d\lambda}\left(
	\lambda \frac{1-\lambda^{N+1}}{1-\lambda}
	\right)\nonumber\\
	&\leq&
	(1-\eta) \frac{1-\eta^{N+1}}{1-\eta}.
\end{eqnarray}
Considering 
\begin{eqnarray}
	f(\lambda,\eta,x):= (1-\eta) \frac{1-\eta^{x+1}}{1-\eta}- (1-\lambda)^2 
	\frac{d}{d\lambda}\left(
	\lambda \frac{1-\lambda^{x+1}}{1-\lambda}
	\right)
\end{eqnarray}
as a function of a real $x\in(1,\infty)$,
it follows immediately from an elementary discussion of the behavior of
$f$ that pairs of $\lambda,\eta\in(0,1)$ with $\eta>\lambda$ can
be found such that $f(\lambda,\eta,x)\geq 0$ for all
$x\in(1,\infty)$ (take, e.g., $\lambda=0.1$, $\eta=0.11$). 
Hence, for such pairs of $\lambda,\eta$,
Ineq.\ (\ref{cr}) is satisfied for all $N=0,1,...$. Therefore, $r,s\in[0,\infty)$
with $r<s$ can be found such that
\begin{equation}
	\rho_r\otimes \rho_r\longrightarrow \rho_s\otimes \rho_0\,\,\text{ under LOCC, but}\,\,\,
	\rho_r\otimes \rho_r\not\longrightarrow \rho_s\otimes \rho_0\,\,\text{ under GLOCC.}
\end{equation}
This argument shows that in principle, by allowing for all LOCC, one
may 'pump' entanglement from two two-mode squeezed states into one of
the two-mode systems, at the expense of reducing the entanglement of
the other system. 

In the case of \Eqref{ex2}, however, even  
with LOCC one can do no more than with GLOCC. To see this, we only
have to look at the sum of squares of the first $K=(N+1)(N+2)/2$
Schmidt coefficients of $\rho_r\otimes \rho_r$ for some $N\in{\mathbbm{N}}$.
We find 
\begin{equation}
L_K := \sum_{k=0}^{K-1} l_k = (1-\lambda)^2 \sum_{k=0}^N(k+1)\lambda^k =
1-\left[1+\frac{N+1}{\cosh^2 (r/2) }\right]\tanh^{2(N+1)}(r/2). 
\end{equation}
This is to be compared with the sum of the first $K$ Schmidt
coefficients of $\rho_s\otimes \rho_0$,
\begin{equation}
M_K  := \sum_{k=0}^{K-1} m_k = 1-\tanh^{2(K+1)}(s/2).
\end{equation}
Noting that $K$ grows quadratically in
$N$, we see that for $N$ large enough $M_{K} > L_{K}$ --- no matter how
large $r>0$. Therefore,
\begin{equation}
	\rho_s \otimes \rho_0 \not\longrightarrow
\rho_r\otimes\rho_r\,\,\,\text {under LOCC} 
\end{equation}
for all $r,s >0$. This transformation is in other words not even possible under
general local operations, and \Eqref{MainCrit} is no further
restriction. This statement is indeed the analogue of the statement
for finite-dimensional systems, the Schmidt number canny increased by LOCC.

In turn, in the asymptotic limit of infinitely many identically
prepared initial states such a dilution procedure becomes possible
again under LOCC for an appropriate choice of $r,s>0$. Then, the
possibility of such a transformation is governed only by the
von-Neumann entropies of the reduced states held by both parties. For
GLOCC, such a dilution of two-mode squeezing stays impossible, even in
the asymptotic limit.

One should keep in mind, however, that the bounds provided by general
LOCC for Gaussian states are extraordinarily optimistic in any
practical context, as general quantum operations are required that act
in infinite-dimensional Hilbert spaces. Such general operations are
certainly beyond all realistic assumptions concerning the set of
feasible operations that are available in actual experiments. This
argument nevertheless points towards the possibility of realizing
non-Gaussian operations that map known Gaussian states onto Gaussian
states. That such maps can have the power to distill quantum
entanglement was shown in \cite{Gaussify} for pure states.

\section{Discussion and Conclusion}

We have presented a general criterion for the possibility of
transforming one pure Gaussian state of $n\times n$ modes into another
by means of Gaussian local operations with classical
communication. This criterion has been put into the context of the
majorization criterion for general local operations with classical
communication.  A very useful generalization would be concerned with a
full criterion for mixed Gaussian quantum states. In fact, some of the
structure of the above proof remains true in the mixed-state case,
however, the normal form of Lemma~\ref{Lstdform} is not
available. This paper can hopefully contribute to paving the way for
finding such a general tool.

\section{Acknowledgments}

GG and JE would like to thank J.~Fiura{\v s}ek and P.~van Loock for
interesting and fruitful discussions on the possibility of applying
the majorization criterion to assess Gaussian state transformations
during the ESF workshop on Continuous-Variable Quantum Information
Processing 2002 in Brussels.  JE and MBP thank K.~Audenaert and
S.~Scheel for discussions on the subject of Gaussian operations.  JE 
would also like to thank J.~Preskill and the IQI group at CalTech for
their kind hospitality. This work has been supported by the European
Union (EQUIP -- IST-1999-11053), the DFG (Schwer\-punkt\-programm
QIV), the A.-v.-Humboldt-Foundation, the NSF (EIA-0086038), and the
ESF, who supported an academic visit of MBP and JE to the
Max-Planck--Institute for Quantum Optics in May 2002, where this work
was initiated.

\section{Appendix A: Proof of Lemma 1}

\smallskip \proof{
We write the CM $\gamma$ of the pure state $\rho$ as
\begin{eqnarray}\label{partit}
    \gamma=\left(
    \begin{array}{cc}
    A & C \\
    C^{T} & B
    \end{array}
    \right).
    \end{eqnarray}
Purity of the state $\rho$ implies that $-(\gamma \sigma)^{2} =
{\mathbbm{1}}$. We first make use of this condition to show that
the symplectic spectrum of $A$ and $B$ are identical. This implies
that local operations allow for achieving 
\begin{equation}
	A=B=\bigoplus_{k} a_{k}
{\mathbbm{1}}_{2n_{k}},
\end{equation} 
where $a_{k}\geq 1$ are the symplectic
eigenvalues of $A$ and $n_{k}$ is their multiplicity. In a second
step we show that without further changing $A$   and
$B$, we can transform $C$   by local Gaussian
operations into $\oplus_{k}(a_{k}^{2}-1)^{1/2} {\mathbbm{1}}_{2
n_{k}}$. Renaming $a_{k}= \cosh (r_{k})$ with an appropriate
$r_{k}\geq 0$, $k=1,...,n$, 
then proves the claim.

(a) All further arguments derive from the equality
\begin{eqnarray}\label{startingp}
    \left(
    \begin{array}{cc}
    A\sigma A\sigma+C\sigma C^{T}\sigma & A\sigma C\sigma+C\sigma B\sigma \\
    C^{T}\sigma A\sigma +B\sigma C^{T}\sigma  & B\sigma B\sigma +C^{T} \sigma C\sigma
    \end{array}
    \right)= -{\mathbbm{1}},
\end{eqnarray}
which holds for all covariance matrices of pure states by virtue
of $-(\gamma \sigma )^{2} = {\mathbbm{1}}$. From the two diagonal
blocks we obtain
\begin{eqnarray}
    A\sigma A\sigma ^{T}= {\mathbbm{1}}+C\sigma C^{T}\sigma ,\,\,
    B\sigma B\sigma ^{T}={\mathbbm{1}}+C^{T}\sigma C\sigma .
\end{eqnarray}
But the spectrum of the matrices on the right hand side is
directly related to the   symplectic spectrum of $A$
and $B$ respectively  : the eigenvalues of $A\sigma
A\sigma ^{T}$ are the squares of the symplectic eigenvalues of
$A$. Since the matrices on the left hand side have the same
spectrum, it follows that $A$ and $B$ have the same symplectic
spectrum. Hence, $A=B= \oplus_k a_{k} {\mathbbm{1}}_{2n_{k}}$, without loss
of generality.

(b) For the second part, observe   from the
off-diagonal blocks in Eq. (\ref{startingp})  
that $A\sigma $ anti-commutes with $C\sigma $ and $C^{T}\sigma $. From
this   and the positivity of the $a_k$   one
shows   directly   that $C$ must be
block-diagonal, with blocks corresponding to the $a_{k}$
eigenspaces of $A$. Now consider Eq.\ (\ref{startingp}) again for
each such block separately. We denote the corresponding $m\times
m$ CM of a pure state as $\gamma'$, partitioned in block form as
in Eq.\ (\ref{partit}). Then $A'=B'= a{\mathbbm{1}}$ for some
$a\geq 1$.   For the off-diagonal block $C'$
  we find 
\begin{equation}
	C'\sigma {C'}^{T}\sigma
=(a^{2}-1){\mathbbm{1}},\,\,\, \sigma {C'}^{T}\sigma
={C'}^{T},
\end{equation}
which imply that $C'{C'}^{T}=
(a^{2}-1){\mathbbm{1}}$, from which it follows that  
$C'=(a^2-1)^{1/2} 
O$, 
 where $O$ is both symplectic and orthogonal and can be removed
 by a local unitary operation without affecting $A'$. Such 
transformations that are both symplectic and orthogonal
correspond to a passive transformation \cite{passive}.
}



\begin{thebibliography}{99}
\bibitem{improper}
	In particular, this holds for the 'eigenstates' of the position operator,
	which are elements of the dual space of Schwartz space.
\bibitem{Niel}
	The validity of the majorisation criterion in 
	this infinite-dimensional setting can be formally derived by
	investigating sequences of finite-dimensional 
	pure states $\{\rho_k\}_{k=0}^\infty$
	and $\{\rho'_k\}_{k=0}^\infty$ defined on $k\times k$-dimensional system 
	such that $\rho_k\rightarrow \rho$
	and $\rho'_k\rightarrow \rho'$ in trace-norm.
	Considering the finite case for all $k\in{\mathbbm{N}}$, one
 	 immediately arrives at the majorisation criterion in the infinite-dimensional case.
\bibitem{passive}
    Note that transformations that are both  symplectic and orthogonal
	reflect
    all those unitary Gaussian operations that can be implemented by means
    of passive optical elements (beam splitters and phase shifts).
\end{thebibliography}
\end{document}